\documentclass[aps,prl,reprint,superscriptaddress,amsmath,amssymb,longbibliography,nobibnotes]{revtex4-2}
\usepackage{iftex}
\ifXeTeX
  \makeatletter
  \input{tuenc.def}
  \makeatother
  \DeclareFontSubstitution{TU}{termesportable}{m}{n}
  \DeclareFontFamily{TU}{termesportable}{}
  \DeclareFontShape{TU}{termesportable}{m}{n}{<-> "[TeXGyreTermesX-Regular.otf]:mapping=tex-text"}{}
  \DeclareFontShape{TU}{termesportable}{m}{it}{<-> "[TeXGyreTermesX-Italic.otf]:mapping=tex-text"}{}
  \DeclareFontShape{TU}{termesportable}{b}{n}{<-> "[TeXGyreTermesX-Bold.otf]:mapping=tex-text"}{}
  \DeclareFontShape{TU}{termesportable}{b}{it}{<-> "[TeXGyreTermesX-BoldItalic.otf]:mapping=tex-text"}{}
  \DeclareFontShape{TU}{termesportable}{bx}{n}{<->ssub*termesportable/b/n}{}
  \DeclareFontShape{TU}{termesportable}{bx}{it}{<->ssub*termesportable/b/it}{}

\else
  \usepackage[T1]{fontenc}
  \usepackage{newtxtext}
\fi
\usepackage{amsthm}
\usepackage{mathtools}

\usepackage[noOT1]{newtxmath}
\DeclareFontShape{OT1}{cmss}{m}{n}{<->cmss10}{}
\usepackage{graphicx}
\usepackage[dvipsnames]{xcolor}
\definecolor{PaperLink}{HTML}{3B5F78}
\usepackage[colorlinks=true,linkcolor=PaperLink,citecolor=PaperLink,urlcolor=PaperLink]{hyperref}
\newcommand{\manuscripttitle}{Fermionic and bosonic synthesis with optimal resource cost}
\hypersetup{pdftitle={\manuscripttitle},pdfsubject={Deterministic synthesis with Gaussian controls}}
\newtheorem{theorem}{Theorem}
\newtheorem{lemma}{Lemma}
\newtheorem{corollary}{Corollary}
\newtheorem{proposition}{Proposition}
\theoremstyle{definition}

\newcommand{\ket}[1]{\lvert #1\rangle}
\newcommand{\bra}[1]{\langle #1\rvert}
\newcommand{\ketbra}[2]{\ket{#1}\bra{#2}}
\newcommand{\braket}[2]{\langle #1\vert #2\rangle}

\newcommand{\mc}{\mathcal}
\newcommand{\mbb}{\mathbb}

\newcommand{\Span}{\operatorname{span}}

\newcommand{\R}{\mbb{R}}

\begin{document}

\title{\manuscripttitle}

\author{Fuchuan Wei}
\thanks{These authors contributed equally to this work.}
\affiliation{Yau Mathematical Sciences Center, Tsinghua University, Beijing 100084, China}

\author{Zhenhuan Liu}
\thanks{These authors contributed equally to this work.}
\affiliation{Quantum Research Center, Technology Innovation Institute (TII), Abu Dhabi, United Arab Emirates}

\author{Zi-Wen Liu}
\email{zwliu0@tsinghua.edu.cn}
\affiliation{Yau Mathematical Sciences Center, Tsinghua University, Beijing 100084, China}

\begin{abstract}

We establish optimal worst-case non-Gaussian gate counts for exact state preparation and unitary synthesis in fermionic and bosonic systems, resolving natural analogues of the extensively studied qubit $T$-count problem where an exponential gap persists for general unitaries. Specifically, with Gaussian unitaries free, we prove optimal costs of $\Theta(\sqrt D)$ for pure state preparation and $\Theta(D)$ for unitary synthesis, where $D=2^n$ for $n$ fermionic modes and $D=2^q$ for $q$ bosonic dual-rail qubits. The cost counts uses of any given non-Gaussian gate that preserves fermion parity or total photon number, respectively.  In both settings, our protocols employ a simple single-particle encoding scheme that maps each logical basis state to a distinct register mode so that passive Gaussian operations implement any target unitary directly, with non-Gaussian gates required only for target-independent encoding and decoding. Matching lower bounds follow from dimension counting and hold for almost every allowed target, even with arbitrarily many ancillary modes, adaptive Gaussian processing, and postselection. Furthermore, we show that coherent circuits obey a mode-gate tradeoff requiring exponential circuit width to reach these minimal gate counts, so our protocols simultaneously achieve optimality in gate cost and ancillary space.

\end{abstract}
\maketitle

\section{Introduction}
\label{sec:introduction}

The power of quantum computing relies on the ability to realize general target states and unitaries, which requires decomposing complex transformations into sequences of directly executable elementary operations. This fundamental problem known as quantum compilation or synthesis has been studied extensively under a wide range of requirements~\cite{barenco1995elementary,vartiainen2004efficient,bergholm2005quantum,dawson2006solovay,shende2006synthesis,shende2009toffoli,plesch2011quantumstatepreparationuniversalgatedecompositions,ross2016optimal,gidney2018halving,zhang2022quantum,sun2023asymptoticallyoptimalcircuitdepthquantumstatepreparationgeneralunitarysynthesis,krol2024beyondshannon,claudon2024polylogarithmic,paradis2024reqomp,low2024trading,khattar2025conditionallyclean,zi2025constantdepthstate,gosset2026quantumstatepreparationoptimaltcount,yuan2026exactunitarysynthesissimultaneouslyoptimalsizedepthancillatradeoff,nehoran2026unitariesconstantdepthquantumcircuits}.

In practice, some elementary operations can be much more resource-intensive to implement than others, so to reduce the overall resource cost, a central goal is to minimize the number of these ``expensive'' operations.
A canonical form is the ``$T$-count'' problem, which asks how many $T$ gates are required in Clifford+$T$ circuit decompositions.
In fault-tolerant architectures, logical Clifford gates are comparatively inexpensive~\cite{gottesman1998theory,bombin2006topological,fowler2012surface}, whereas logical $T$ gates typically constitute the dominant bottleneck in spacetime cost~\cite{zhou2000methodology,bravyikitaev2005magic,eastin2009restrictions,bravyi2013classification,litinski2019game,beverland2020nonclifford,gidney2024cultivation}.
Beyond fault tolerance overhead, the $T$-count also governs classical simulation complexity, since Clifford circuits can be simulated efficiently~\cite{aaronson2004stabilizer} while known algorithms for Clifford+$T$ circuits have runtimes exponential in the number of $T$ gates~\cite{bravyi2016classical,bravyi2019simulation}.
Optimal or near-optimal $T$-count bounds are known for structured targets such as multi-qubit Toffoli gates~\cite{jones2013low,gidney2018halving,beverland2020nonclifford,gosset2025multiqubittoffoli}, near-Clifford unitaries~\cite{fang2026unitarysynthesisnearoptimaltcountnearcliffordunitaries}, and diagonal unitaries~\cite{gosset2026quantumstatepreparationoptimaltcount}.
At constant approximation error and with unrestricted ancillary qubits, a tight worst-case $T$-count of $\Theta(2^{n/2})$ has also  been established for arbitrary $n$-qubit pure state preparation~\cite{gosset2026quantumstatepreparationoptimaltcount}.
By contrast, the optimal worst-case $T$-count for general $n$-qubit unitary synthesis has remained a longstanding open question despite sustained effort and a recent surge of progress.
Up to polynomial factors, the upper bound has been successively improved from $2^{2n}$ via gate-by-gate compilation~\cite{barenco1995elementary,knill1995approximation,vartiainen2004efficient,mottonen2004quantum,shende2006synthesis,dawson2006solovay,kliuchnikov2013exact,selinger2015efficient,ross2016optimal} to $2^{3n/2}$~\cite{low2024trading}, $2^{4n/3}$~\cite{tan2026unitarysynthesisfewertgates}, $2^{5n/4}$~\cite{yuan2026blockflattening}, and most recently $2^{8n/7}$~\cite{liu2026tefficientunitarysynthesisblocksparsehamiltoniansimulation}, yet an exponential gap remains above the known $\Omega(2^n)$ lower bound~\cite{gosset2026quantumstatepreparationoptimaltcount,low2024trading}.

\begin{figure}[tphb]
\centering
\includegraphics[width=\columnwidth]{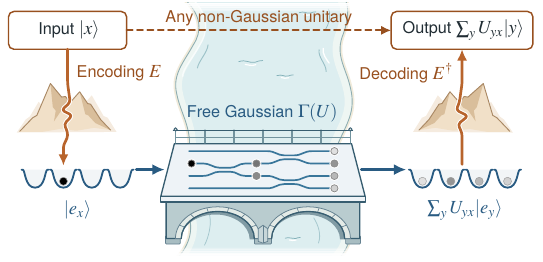}
\caption{Optimal fermionic and bosonic unitary synthesis via single-particle encoding.
The dashed arrow across the river shows the target action of an arbitrary non-Gaussian unitary $U$ on the data register.
We encode each input basis state $\ket{x}$ as a single particle in the corresponding mode $\ket{e_x}$.
The interferometer on the bridge implements the target action on the encoded system as a free Gaussian unitary $\Gamma(U)$.
The two arrows over the mountains represent encoding $E$ and decoding $E^\dagger$, which together carry the entire non-Gaussian resource cost.}
\label{figure:single-particle-encoding}
\end{figure}

{Beyond qubit architectures, quantum information processing directly with fermionic and bosonic degrees of freedom is of fundamental interest across diverse physical settings: fermionic modes naturally describe topological Majorana platforms, ultracold atomic lattices, and native fermionic simulation without nonlocal encoding overhead~\cite{kitaev2001unpairedmajorana,bravyi2002fermionicquantumcomputation,nayak2008nonabelian,gonzalezcuadra2023fermionic,bojovic2026collisional}, while bosonic modes underpin photonic linear-optical, continuous-variable, and bosonic code architectures~\cite{knill2000efficientlinearopticsquantum,gottesman2001encoding,kok2006reviewarticlelinearoptical,weedbrook2012gaussian,joshi2021quantum,hackl2021bosonicfermionicgaussianstateskahlerstructures}.
In both settings, Gaussian unitaries, generated by quadratic Hamiltonians, correspond to readily accessible operations such as single-particle tunneling in atomic systems and mode mixing or squeezing via beam splitters, phase shifters, and squeezers in optics~\cite{reck1994unitary,knill2001fermioniclinearopticsmatchgates,terhal2002noninteractingfermion,braunstein2005squeezing,weedbrook2012gaussian,clements2016optimal,bojovic2026collisional}.
By contrast, coherent non-Gaussian gates require interactions beyond quadratic order and demand substantially finer control, as in engineered collisions between fermionic atoms or optical nonlinearities and cavity-mediated interactions between photons~\cite{milburn1989fredkin,vitelli2013joining,passaro2013joiningsplittingquantumstatesphotons,hacker2016photon,gonzalezcuadra2023fermionic,bojovic2026collisional}. Therefore, non-Gaussian operations parallel non-Clifford gates in qubit systems, serving as an essential quantum computational resource~\cite{albarelli2018resource,takagi2018convex,hebenstreit2019purefermionicnongaussianstates,chabaud2020stellar,liu2022manybodyquantummagic,tarabunga2026fermionicnongaussianitybellsamplingmonotonesefficientquantumalgorithms,turkeshi2026williamsonmajorizationtheoryfermionicnongaussianity,wei2026robustexponentiallowerboundsfermionicbosonicgaussianranks,reardonsmith2026exponentiallowerboundsfermionicgaussianrankmagicstatesbosoniccoherentstaterankfockstates}.
Also mirroring the qubit case, Gaussian circuits with Gaussian inputs and measurements can be simulated efficiently~\cite{knill2001fermioniclinearopticsmatchgates,terhal2002noninteractingfermion,bartlett2002efficient,bravyi2004lagrangianrepresentationfermioniclinear}, whereas known classical algorithms scale exponentially in runtime with the number of non-Gaussian gates~\cite{chabaud2021classical,dias2024classicalsimulationnongaussianfermionic,reardonsmith2024improvedsimulationquantumcircuitsdominatedfreefermionicoperations}.
Furthermore, non-Gaussian operations are indispensable even for entanglement distillation and quantum error correction~\cite{eisert2002distilling,giedke2002characterizationgaussianoperationsdistillation,niset2009nogotheoremgaussianquantumerrorcorrection,tang2026fermionicquantumerrorcorrectionneverfree}, and the number of non-Gaussian gates governs the complexity of learning fermionic and bosonic systems~\cite{austin2025efficientlylearningfermionicunitariesfewnongaussiangates,mele2025learningcv}.
These considerations directly motivate minimizing the non-Gaussian gate count for state preparation and unitary synthesis, yet prior results are mostly limited to universality criteria and explicit decompositions without optimality guarantees~\cite{lloyd1999quantum,bravyi2002fermionicquantumcomputation,sefi2011decompose,passaro2013joiningsplittingquantumstatesphotons,oszmaniec2017universalextensionsrestrictedclassesquantumoperations,hebenstreit2019purefermionicnongaussianstates,arrazola2022universal}, leaving the optimal cost largely uncharted.}

In this work, we resolve this question for both fermionic and bosonic systems by establishing the optimal worst-case non-Gaussian gate costs for state preparation and unitary synthesis through a unified mathematical structure.
With Gaussian operations free and allowing ancillary modes, we prove that the optimal worst-case non-Gaussian gate counts are tightly $\Theta(\sqrt{D})$ for pure state preparation and $\Theta(D)$ for unitary synthesis, where $D = 2^n$ for $n$ fermionic modes and $D = 2^q$ for $q$ bosonic dual-rail qubits, counting uses of a given bounded-support non-Gaussian gate that preserves fermion parity or total photon number, respectively.
In contrast to the familiar qubit setting where an exponential gap remains, our results close the gap completely up to constant factors.

In both fermionic and bosonic settings, our upper bounds stem from a simple unified construction that leverages a basic property of passive Gaussian unitaries, namely they can implement arbitrary unitaries on the single-particle subspace, as illustrated in Fig.~\ref{figure:single-particle-encoding}.
By encoding computational basis states as single-particle excitations in distinct modes, we implement the target unitary by free Gaussian interferometry, with non-Gaussian gates needed only for target-independent encoding and decoding.
The matching lower bounds follow from dimension-counting arguments to bound the effective degrees of freedom accessible at a given non-Gaussian gate count.
For coherent circuits, this counting argument further shows that almost every target requires exponentially many ancillary modes to attain the optimal gate count, with our constructions also achieving the optimal mode scaling for both tasks.

\section{Settings and main results}
\label{section:settings-and-main-results}

In this section, we overview the problem setting, the main results, and the techniques to prove the upper and lower bound of the compilation costs.
We consider fermionic and bosonic synthesis with free Gaussian ancillary states and Gaussian unitaries, both of which can be used without limit.
Their definitions and properties are reviewed in Supplemental Material Sec.~\ref{appendix:gaussian-models}.
With access to continuously tunable Gaussian controls, we study exact synthesis.
Let $W$ be a non-Gaussian gate supported on a bounded number of modes.
For each target, we minimize the number of uses of $W$ needed to prepare the state or implement the unitary exactly.

In the fermionic model, we prepare a pure state of definite parity from a fixed occupation state in the same parity sector and implement parity-preserving unitaries on the full Fock space.
In the bosonic model, we supply one photon per dual-rail qubit, with logical basis states $\ket{0}_{\rm L}=\ket{10}$ and $\ket{1}_{\rm L}=\ket{01}$~\cite{knill2000efficientlinearopticsquantum,kok2006reviewarticlelinearoptical}. For $q$ qubits, we prepare states from $\ket{0}_{\rm L}^{\otimes q}$ and synthesize unitaries on the full $2^q$-dimensional space.
The main results of this work can be summarized with the following theorem.

\begin{theorem}[Optimal synthesis]
\label{theorem:main-overview}
Let $D=2^n$ for $n$ fermionic modes or $D=2^q$ for $q$ dual-rail qubits.
Let $W$ be any non-Gaussian gate that acts on a bounded number of modes and preserves fermion parity or total photon number, respectively.
The worst-case minimum number of uses of $W$ for exact synthesis is
\begin{align}
&\Theta(\sqrt D)\ \quad \text{for pure state preparation},\\ &\Theta(D)\  \quad\text{for unitary synthesis}.
\end{align}
The lower bounds hold for almost every allowed target, even with an arbitrary finite number of ancillary modes, adaptive Gaussian processing, and postselection.
\end{theorem}

In both models, single-particle encoding lets us apply the target matrix $U$ with a Gaussian unitary $\Gamma(U)$, see Fig.~\ref{figure:single-particle-encoding}.
With one mode per input basis state, the single-particle register has $2^n$ modes for $n$ fermionic input modes and $2^q$ modes for $q$ dual-rail qubits.
For input $\ket{x}$, the encoder $E$ prepares the register in $\ket{e_x}$, the state with one particle in mode $x$.
Every unitary matrix $U$ can be written as $e^{-ih}$ with $h$ Hermitian.
Let $a_x$ be the annihilation operator for mode labeled by $x$, so that $\ket{e_x}=a_x^\dagger\ket{0}$.
Consider the number-preserving quadratic Hamiltonian $H(h)=\sum_{x,y} h_{yx}a_y^\dagger a_x.$
Since $a_y^\dagger a_x\ket{e_z}=\delta_{xz}\ket{e_y}$, the matrix of $H(h)$ on the single-particle subspace is exactly $h$ in the label basis.
The passive Gaussian unitary $\Gamma(U)=e^{-iH(h)}$ therefore implements $U=e^{-ih}$ on that subspace:
\begin{equation}
\Gamma(U)\ket{e_x}=\sum_y U_{yx}\ket{e_y}.
\end{equation}
Decoding maps the output labels back to the data basis, giving $U\ket{x}$ and restoring the ancillary inputs.
Thus, all non-Gaussian resources in this protocol are spent on encoding and decoding, both of which are independent of the target unitary $U$.
Together, the encoder and decoder require $O(D)$ uses of $W$, as detailed in Section~\ref{sec:single-particle-encoding}.

The cost for preparing a state can be further reduced using Schmidt decomposition.
For a target state $\ket{\psi}$, we divide the system as evenly as possible into blocks $A$ and $B$.
Let $r_\psi=O(\sqrt{D})$ be its Schmidt rank across this partition and write $\ket{\psi}=\sum_{j=0}^{r_\psi-1}\lambda_j\ket{u_j}_A\ket{v_j}_B$.
In the single-particle register, we represent Schmidt label $j$ by the state $\ket{e_j}$ of one particle in mode $j$.
We first prepare $\sum_{j=0}^{r_\psi-1}\lambda_j\ket{e_j}_A\ket{e_0}_B$ by superposing the modes in $A$ while leaving the particle in $B$ in mode $0$.
For each nonzero label $j$, we swap modes $0$ and $j$ in $B$ when mode $j$ in $A$ is occupied.
Together these swaps copy the label from $A$ to $B$, producing $\sum_j \lambda_j\ket{e_j}_A\ket{e_j}_B$.
We rotate each register so that the amplitudes of its labels match the occupation-basis coefficients of the corresponding Schmidt vectors $\ket{u_j}_A$ and $\ket{v_j}_B$. The two block decoders then produce $\ket{\psi}$.
In addition to  decoding, the controlled swaps that correlate the Schmidt labels require non-Gaussian gates. 
Since both $A$ and $B$ have dimension $O(\sqrt D)$, the two decoders and the controlled swaps each require $O(\sqrt D)$ non-Gaussian gates in total.

\begin{figure}
    \centering
    \includegraphics[width=0.98\linewidth]{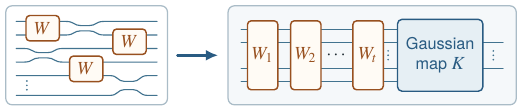}
    \caption{Lower bounds from parameter counting.
    The circuit on the left shows a successful branch with $t$ uses of the non-Gaussian gate $W$. Each illustrated use of $W$ involves two modes.
    Moving the Gaussian unitaries to the end conjugates each instance of $W$, giving $W_1,\cdots,W_t$ on a compressed set of modes.
    The final Gaussian unitary and deferred measurement outcomes define the conditional Gaussian map $K$.}
    \label{figure:lower-bound}
\end{figure}

Our lower bounds follow from a simple parameter-counting argument that also applies to circuits with intermediate measurements and postselection.
On a successful branch with $t$ uses of $W$, we fix the measurement record and separate the Gaussian operations from the non-Gaussian ones, as illustrated in Fig.~\ref{figure:lower-bound}.
This separation lets us bound the number of independent parameters that can affect the branch output.

After deferring the measurements, we move the Gaussian unitaries to the end of the circuit, conjugating each instance of $W$ by the preceding Gaussian evolution.
Let $m$ denote the number of data modes, with $m=n$ for fermions and $m=2q$ for bosons.
The input data modes and the transformed mode directions of the conjugated gates span a register of $m_{\mathrm{eff}}\le m+O(t)$ effective modes, since each use of $W$ involves a bounded number of modes.
The final Gaussian unitary, the remaining vacuum inputs, and the deferred measurement outcomes define a conditional Gaussian map $K$ from the effective register to the $m$ output modes.
Each conjugated gate $W_j$ is determined by the images of the mode operators on the support of $W$, which require $O(m_{\mathrm{eff}})$ real parameters in total.
The Gaussian kernel of $K$ requires $O((m_{\mathrm{eff}}+m)^2)$ real parameters, so the $t$ conjugated gates and $K$ together require $O((m+t)^2)$ parameters.

In both the fermionic and bosonic settings, the target manifolds have real dimensions $\Theta(D)$ for pure states and $\Theta(D^2)$ for unitaries.
Since $m=O(\log D)$ in both models, comparing these dimensions with the parameter count gives the lower bounds of $\Omega(\sqrt D)$ uses of $W$ for almost every state and $\Omega(D)$ uses of $W$ for almost every unitary.
For detailed proofs, we refer to Theorems~\ref{theorem:fermionic-lower} and~\ref{theorem:bos-code-lower} in the Supplemental Material.

\section{Single-particle encoding scheme}
\label{sec:single-particle-encoding}
\label{sec:fermionic-synthesis}
\label{sec:bosonic-synthesis}

We now construct the single-particle encoder $E$ of Fig.~\ref{figure:single-particle-encoding} for inputs on $n$ fermionic modes ($D=2^n$) or $q$ dual-rail qubits ($D=2^q$).
To obtain the upper bound in Theorem~\ref{theorem:main-overview}, it suffices to show that encoding and decoding each require $O(D)$ uses of $W$.

\noindent\textbf{Fermions.}
Both the Gaussian unitaries and the non-Gaussian gate $W$ preserve total parity.
The single-particle register has odd parity before and after encoding, so the input parity must be retained outside this register.
Let $\ket{x}=\ket{x_1,\cdots,x_n}$ denote an input basis state, where $x_j\in\{0,1\}$ is the occupation of data mode $j$.
With a separate parity mode $P$ initially in vacuum, the encoder acts as
\begin{equation}
E_n\bigl(\ket{0}_P\ket{x}\ket{e_{0^n}}\bigr)=\ket{x_1\oplus\cdots\oplus x_n}_P\ket{0^n}\ket{e_x}.
\label{equation:main-encoding}
\end{equation}

The encoder $E_n$ transfers the input string $x$ to the single-particle register one bit at a time.
For $j=1,\cdots,n$, it first records $x_j$ and then clears data mode $j$ to $\ket{0}$ while updating the parity stored in $P$.
To record $x_j$ in the single-particle register, the encoder pairs the modes labeled $z0\,0^{n-j}$ and $z1\,0^{n-j}$ for every prefix $z$ of length $j-1$.
$x_j$ controls a fermionic swap on each pair.
The fermionic swap exchanges $\ket{01}$ and $\ket{10}$ while mapping $\ket{11}$ to $-\ket{11}$.
When $x_j=1$, its action on the single-particle register is $\ket{e_{z0\,0^{n-j}}}\leftrightarrow\ket{e_{z1\,0^{n-j}}}$.
For input $\ket{x}$, the state of the single-particle register after the swaps at step $j$ is $\ket{e_{x_1\cdots x_j 0^{n-j}}}$.

We then coherently return data mode $j$ to vacuum, disentangling it from the single-particle register.
Clearing an occupied mode changes the data parity, so we flip $P$ when $x_j=1$ to preserve total parity.
For each prefix $z$, the occupation of register mode $z1\,0^{n-j}$ controls a pair flip $Q_j$ on $P$ and data mode $j$.
For the Jordan--Wigner ordering beginning with $P,1,\cdots,n$, the pair flip (Gaussian) is $Q_j=-i\gamma_2\gamma_{2j+1}=X_P\bigl(\prod_{k=1}^{j-1}Z_k\bigr)X_j$, where $X_k$ and $Z_k$ are Pauli operators on data mode $k$.
Since the first $j-1$ data modes are already in $\ket{0^{j-1}}$, the $Z$ string has eigenvalue $+1$ and $Q_j$ acts as $X_P X_j$.
For a basis input $\ket{x}$, the joint state after step $j$ is $\ket{x_1\oplus\cdots\oplus x_j}_P\ket{0^j x_{j+1}\cdots x_n}\ket{e_{x_1\cdots x_j0^{n-j}}}$.

After $n$ steps, the single-particle register is in $\ket{e_x}$, all data modes are empty, and $P$ holds the input parity, as in Eq.~\eqref{equation:main-encoding}.
Step $j$ uses one controlled fermionic swap and one controlled pair flip, which are all three-mode non-Gaussian gates, for each of the $2^{j-1}$ mode pairs, giving $2^n-1$ gates of each type over all $n$ steps.

The fermionic swap and the pair flip $Q_j$ are both two-mode Gaussian unitaries.
By Proposition~\ref{proposition:controlled-fermionic-gaussian}, Gaussian unitaries and $O(1)$ uses of $W$ suffice to control any parity-preserving Gaussian unitary on a bounded number of target modes.
To implement this control, we decompose the target into rotations in Majorana planes and use opposite half rotations $\pm\theta/2$ that cancel for an empty control mode.
A conditional phase gate synthesized from $W$ and Gaussian unitaries acts before and after the second rotation, reversing it for an occupied control mode to give the full angle $\theta$.

Since the controlled fermionic swaps and controlled pair flips are self-inverse, the decoder applies the same $2(2^n-1)$ gates in reverse order.
Encoding and decoding therefore each require $O(2^n)$ uses of $W$.

\noindent\textbf{Bosons.}
Recall that the dual-rail basis states are $\ket{0}_{\rm L}=\ket{10}$ and $\ket{1}_{\rm L}=\ket{01}$.
We write $\ket{\mathrm{vac}}_{\rm L}\coloneqq\ket{00}$ for an empty rail pair.
Since passive linear optics (Gaussian operations implemented with standard beam splitters and phase shifters) and $W$ preserve photon number, the encoder moves one of the $q$ input photons into the single-particle register.
To keep this register disentangled from the remaining photons for arbitrary input superpositions, the encoder leaves the other $q-1$ photons in the same state $\ket{0^{q-1}}_{\rm L}$ for every $x$.
With the $2^q$ modes of the single-particle register initially in vacuum, the encoder acts as
\begin{equation}
E_q\bigl(\ket{x}_{\rm L}\ket{0^{2^q}}\bigr)=\ket{\mathrm{vac}\,0^{q-1}}_{\rm L}\ket{e_x}.
\label{equation:main-boson-encoder}
\end{equation}

To supply the single-particle register with a photon, the encoder swaps the two rails of qubit $1$ with register modes labeled by $0^q$ and $10^{q-1}$, respectively.
These swaps transfer the photon to mode $x_1 0^{q-1}$ and leave the first rail pair empty.
For $j=2,\cdots,q$, the encoder records $x_j$ in the single-particle register and then resets qubit $j$ to $\ket{0}_{\rm L}$.
This transfers the state of qubit $j$ to the single photon already carrying the first $j-1$ qubits, a process called photonic state joining~\cite{vitelli2013joining,passaro2013joiningsplittingquantumstatesphotons}.

For $j=2,\cdots,q$, the encoder pairs the register modes labeled $z0\,0^{q-j}$ and $z1\,0^{q-j}$ for each prefix $z$ of length $j-1$.
The second rail of qubit $j$ controls a swap within each pair.
To reset data qubit $j$, the encoder uses each mode labeled $z1\,0^{q-j}$ to control a swap of the two rails of that qubit.
For each of the $2^{j-1}$ mode pairs at step $j$, one controlled swap records $x_j$ and another resets qubit $j$.
Summing over $j=2,\cdots,q$ gives $2(2^q-2)$ controlled swaps.
The decoder applies these swaps in reverse order and then undoes the initial passive mode permutation.

For sufficiently large $q$, each controlled swap requires only passive optics and $O(1)$ uses of $W$ (Corollary~\ref{corollary:bos-controlled-swap}), so encoding and decoding each require $O(2^q)$ uses of $W$.
An exchange is needed precisely when the control mode and the target pair each contain one photon.
We borrow photons from other dual-rail qubits to bring these inputs into a sector where $W$ and passive optics are universal~\cite{oszmaniec2017universalextensionsrestrictedclassesquantumoperations}.
By Lemma~\ref{lemma:bos-fixed-sector}, we can implement this exchange exactly while preserving the state of the borrowed qubits.
For all other allowed inputs, the participating modes contain fewer photons and the circuit acts as the identity.

\section{Ancillary space and gate cost}
\label{sec:mode-gate-tradeoff}

The constructions above attain the optimal non-Gaussian gate counts in both models, using exponentially many ancillary modes. Is it possible to reduce this ancillary space while retaining the same gate count scaling? The following tradeoff shows that the exponential width is necessary for almost every target.

\begin{theorem}[Mode--gate tradeoff]
\label{theorem:mode-gate-tradeoff}
Consider either synthesis model with non-Gaussian gate $W$ as in Theorem~\ref{theorem:main-overview}.
Let $m_{\mathrm{tot}}$ be the total number of data and ancillary modes. For each sufficiently large $D$ and almost every allowed target, the number of uses of $W$ for exact coherent synthesis is
\begin{equation}
\label{equation:mode-gate-tradeoff}
\begin{aligned}
&\Omega\bigl(\max\{\sqrt D,D/m_{\mathrm{tot}}\}\bigr) \quad \text{for pure state preparation},\\
&\Omega\bigl(\max\{D,D^2/m_{\mathrm{tot}}\}\bigr) \quad \text{for unitary synthesis}.
\end{aligned}
\end{equation}
\end{theorem}

As in the proof of Theorem~\ref{theorem:main-overview}, we move all Gaussian unitaries to the end of the circuit. Each conjugated gate $W_j$ contributes at most $O(m_{\mathrm{tot}})$ real parameters, and the final Gaussian unitary contributes at most $O(m_{\mathrm{tot}}^2)$. A circuit with $t$ uses of $W$ therefore has at most $O(m_{\mathrm{tot}}^2+tm_{\mathrm{tot}})$ effective parameters. If $m_{\mathrm{tot}}\le t$, this count is $O(tm_{\mathrm{tot}})$. Comparing it with the target dimensions $\Theta(D)$ for pure states and $\Theta(D^2)$ for unitaries gives the lower bounds $\Omega(D/m_{\mathrm{tot}})$ and $\Omega(D^2/m_{\mathrm{tot}})$, respectively. If $m_{\mathrm{tot}}>t$, then $tm_{\mathrm{tot}}>t^2$, so the width-independent lower bounds $\Omega(\sqrt D)$ and $\Omega(D)$ already imply the corresponding width-dependent bounds. Combining the two bounds for each task gives \eqref{equation:mode-gate-tradeoff}.

For almost every target, achieving the optimal gate-count scaling of $O(\sqrt D)$ for pure state preparation or $O(D)$ for unitary synthesis requires $\Omega(\sqrt D)$ or $\Omega(D)$ ancillary modes, respectively, since the data register occupies only logarithmically many modes. Our single-particle encoding and decoding constructions match these mode counts and thus achieve asymptotic optimality in both non-Gaussian gate count and ancillary space.

\section{Discussion}

We have determined the optimal worst-case non-Gaussian costs of state preparation and unitary synthesis in fermionic and bosonic systems. For both tasks, the optimal non-Gaussian gate count scales as the square root of the number of independent real parameters needed to specify a generic target. This scaling reflects a fundamental property of Gaussianity, namely that Gaussian states and processes admit matrix descriptions of mode correlations and their transformations, with a number of independent parameters that grows quadratically with the number of modes. In our setting, the relevant number of modes is proportional to the non-Gaussian gate count, so the number of parameters in this description grows quadratically with that count.

Several questions are worth further investigation.
A natural extension is to ask how the non-Gaussian gate count scales with Kraus rank for general channel synthesis and with state rank for mixed state preparation, and whether these costs can exceed those of unitary synthesis and pure state preparation, respectively.
A second extension is to study approximate compilation. We expect the worst-case gate-count scaling for arbitrary pure states and unitaries to persist at small constant error. For structured gates, allowing approximation may substantially reduce the cost relative to exact synthesis~\cite{gosset2025multiqubittoffoli}.
Even for exact compilation, tailoring the circuit to the target may save resources, raising the question of whether structured targets such as logical states or encoding isometries of quantum codes can be synthesized with fewer non-Gaussian gates.

\section*{\acknowledgmentsname}
We thank Kong-Wing Wu for valuable discussion.
F.W.\ and Z.-W.L.\ are supported in part by NSFC under Grant No.~12475023, Dushi Program, and a startup funding from YMSC.
F.W.\ also acknowledges support from the Shuimu Tsinghua Scholar Program.


%

\clearpage
\onecolumngrid
\setcounter{page}{1}
\renewcommand{\thepage}{S\arabic{page}}
\raggedbottom
\setcounter{equation}{0}
\renewcommand{\theequation}{S\arabic{equation}}
\renewcommand{\theHequation}{supp.\arabic{equation}}
\setcounter{section}{0}
\setcounter{secnumdepth}{2}
\renewcommand{\thesection}{S\arabic{section}}
\renewcommand{\theHsection}{supp.\arabic{section}}
\setcounter{subsection}{0}
\setcounter{theorem}{0}
\setcounter{lemma}{0}
\setcounter{proposition}{0}
\setcounter{corollary}{0}
\setcounter{definition}{0}
\renewcommand{\thetheorem}{S\arabic{theorem}}
\renewcommand{\theHtheorem}{supp.\arabic{theorem}}
\renewcommand{\thelemma}{S\arabic{lemma}}
\renewcommand{\theHlemma}{supp.\arabic{lemma}}
\renewcommand{\theproposition}{S\arabic{proposition}}
\renewcommand{\theHproposition}{supp.\arabic{proposition}}
\renewcommand{\thecorollary}{S\arabic{corollary}}
\renewcommand{\theHcorollary}{supp.\arabic{corollary}}
\renewcommand{\thedefinition}{S\arabic{definition}}
\renewcommand{\theHdefinition}{supp.\arabic{definition}}
\begin{center}
{\large\bfseries Supplemental Material:\par}
\vspace{3pt}
{\large\bfseries Fermionic and bosonic synthesis with optimal resource cost\par}
\end{center}
\newwrite\SMtocfile
\newcommand{\SMsection}[2]{%
  \section{#1}\label{#2}%
  \immediate\write\SMtocfile{\string\SMtocsection{\unexpanded{#1}}{#2}}%
}
\newcommand{\SMsubsection}[2]{%
  \subsection{#1}\label{#2}%
  \immediate\write\SMtocfile{\string\SMtocsubsection{\unexpanded{#1}}{#2}}%
}
\newcommand{\SMtocline}[4]{%
  \par\noindent\hspace*{#1}%
  \hyperref[#4]{{#2\ref*{#4}.\enspace #3}}%
  \nobreak\dotfill\pageref{#4}\par
}
\newcommand{\SMtocsection}[2]{%
  \addvspace{2pt}\SMtocline{0pt}{\bfseries}{#1}{#2}%
}
\newcommand{\SMtocsubsection}[2]{\SMtocline{1.7em}{}{#1}{#2}}

\begin{center}
  \textbf{Contents}
\end{center}
\InputIfFileExists{\jobname.smtoc}{}{}%
\immediate\openout\SMtocfile=\jobname.smtoc\relax
\par\bigskip

\SMsection{Gaussian states and unitaries}{appendix:gaussian-models}

We review Gaussian states and unitaries in fermionic and bosonic systems.
We describe how Gaussian unitaries transform fermionic Majorana operators and bosonic creation and annihilation operators, and give the parameter counts used to prove our lower bounds.

\SMsubsection{Fermionic Gaussian operations}{appendix:fermionic-model}

For $m$ fermionic modes, let $a_j^\dagger$ and $a_j$ denote the creation and annihilation operators of mode $j$.
They satisfy the canonical anticommutation relations $\{a_i,a_j^\dagger\}=\delta_{ij}I$ and $\{a_i,a_j\}=\{a_i^\dagger,a_j^\dagger\}=0$.
The two Majorana operators of mode $j$ are~\cite{kitaev2001unpairedmajorana,bravyi2002fermionicquantumcomputation}
\begin{equation}
\gamma_{2j-1}\coloneqq a_j+a_j^\dagger,\quad \gamma_{2j}\coloneqq i(a_j^\dagger-a_j).
\label{equation:prelim-majorana-mode}
\end{equation}
They obey $\{\gamma_\mu,\gamma_\nu\}=2\delta_{\mu\nu}I$.
Up to an overall phase, fermionic Gaussian unitaries are generated by quadratic Majorana Hamiltonians~\cite{hackl2021bosonicfermionicgaussianstateskahlerstructures},
\begin{equation}
H_J=\frac{i}{4}\sum_{\mu,\nu=1}^{2m}J_{\mu\nu}\gamma_\mu\gamma_\nu,\quad J^T=-J,\quad J\text{ real}.
\end{equation}
For $G=e^{-iH_J}$,
\begin{equation}
G^\dagger\gamma_\mu G=\sum_{\nu=1}^{2m}R_{\mu\nu}\gamma_\nu,\quad R=e^J\in\mathrm{SO}(2m).
\label{equation:prelim-fermionic-majorana-action}
\end{equation}
Conversely, every $R\in\mathrm{SO}(2m)$ arises as the Majorana action of a fermionic Gaussian unitary.
Fermionic Gaussian unitaries preserve fermion parity~\cite{terhal2002noninteractingfermion,knill2001fermioniclinearopticsmatchgates,bravyi2004lagrangianrepresentationfermioniclinear}.
Since the Majorana action determines $G$ up to phase, a Gaussian unitary has $\dim\mathrm{SO}(2m)=m(2m-1)$ real parameters.
Pure fermionic Gaussian states are obtained by applying Gaussian unitaries to occupation states.
In each parity sector, their rays form an orbit isomorphic to $\mathrm{SO}(2m)/\mathrm{U}(m)$, of real dimension $m(m-1)$~\cite{hackl2021bosonicfermionicgaussianstateskahlerstructures}.
Up to an overall scalar, a conditional fermionic Gaussian map $K$ from $m_{\mathrm{eff}}$ input to $m$ output modes is determined by the Majorana covariance matrix of its pure Gaussian Choi state.
This follows by applying $K$ to the input half of a full-Schmidt-rank Gaussian Bell state and normalizing the resulting vector.
The pure state covariance therefore has $(m_{\mathrm{eff}}+m)(m_{\mathrm{eff}}+m-1)$ independent real parameters, by the dimension formula above.

To implement a mode rotation $U\in\mathrm{U}(m)$, write $U=e^{-ih}$ with $h=h^\dagger$ and set $H(h)\coloneqq\sum_{i,j=1}^m h_{ij}a_i^\dagger a_j$.
The Gaussian unitary $\Gamma(U)=e^{-iH(h)}$ conserves particle number, fixes the vacuum, and satisfies $\Gamma(U)a_j^\dagger\Gamma(U)^\dagger=\sum_{i=1}^m U_{ij}a_i^\dagger$.
Thus on the single-particle states $\ket{e_j}=a_j^\dagger\ket{\mathrm{vac}}$ we have
\begin{equation}
\Gamma(U)\ket{e_j}=\sum_{i=1}^m U_{ij}\ket{e_i}.
\label{equation:prelim-one-particle}
\end{equation}

\SMsubsection{Bosonic Gaussian operations}{appendix:bosonic-model}

For $m$ bosonic modes, let $b_j^\dagger$ and $b_j$ denote the creation and annihilation operators of mode $j$.
They satisfy the canonical commutation relations $[b_i,b_j^\dagger]=\delta_{ij}I$ and $[b_i,b_j]=[b_i^\dagger,b_j^\dagger]=0$.
On occupation states, $b^\dagger\ket{s}=\sqrt{s+1}\ket{s+1}$ for $s\ge0$ and $b\ket{s}=\sqrt{s}\ket{s-1}$ for $s\ge1$, with $b\ket{0}=0$.
The two quadratures of mode $j$ are $q_j\coloneqq(b_j+b_j^\dagger)/\sqrt{2}$ and $p_j\coloneqq i(b_j^\dagger-b_j)/\sqrt{2}$.
For $\boldsymbol r\coloneqq(q_1,p_1,\cdots,q_m,p_m)^T$, the commutators are $[r_\mu,r_\nu]=i\Omega_{\mu\nu}I$, where $\Omega$ is the standard symplectic form in this ordering.

Bosonic Gaussian unitaries are generated by Hamiltonians of degree at most two in the quadratures, with linear terms generating displacements~\cite{weedbrook2012gaussian}.
For a bosonic Gaussian unitary $G$, the quadrature action is $G^\dagger\boldsymbol r G=S\boldsymbol r+\boldsymbol s I$, where $S$ is real and satisfies $S\Omega S^T=\Omega$ and $\boldsymbol s\in\mbb{R}^{2m}$.
Conversely, every such pair $(S,\boldsymbol s)$ arises from a bosonic Gaussian unitary.
Since the quadrature action determines $G$ up to phase, a Gaussian unitary has $\dim\mathrm{Sp}(2m,\mbb{R})+2m=m(2m+1)+2m$ real parameters.

Conjugating the creation operator \(b_j^\dagger\) by \(G\) gives
\begin{equation}
G b_j^\dagger G^\dagger=d_j I+\sum_{i=1}^m(X_{ij}b_i^\dagger+Y_{ij}b_i).
\label{equation:optical-gaussian-action}
\end{equation}
The coefficients $d_j$ describe displacement, and the $Y_{ij}$ describe the mixing of creation and annihilation operators under squeezing.

Pure bosonic Gaussian states are obtained by applying Gaussian unitaries to the vacuum.
A conditional bosonic Gaussian map $K$ from $m_{\mathrm{eff}}$ input to $m$ output modes is determined by its position-space kernel.
By Lemma~\ref{lemma:bos-gaussian-kernels}, the kernel can be described by $O((m_{\mathrm{eff}}+m)^2)$ real parameters, independently of the number of internal ancillas and measurements.

For the single-photon register used in the upper bounds, take a mode rotation $U\in\mathrm{U}(m)$, write $U=e^{-ih}$ with $h=h^\dagger$, and set $H(h)\coloneqq\sum_{i,j=1}^m h_{ij}b_i^\dagger b_j$.
The Gaussian unitary $\Gamma(U)=e^{-iH(h)}$ conserves photon number and fixes the vacuum.
It can be implemented with beam splitters and phase shifters~\cite{reck1994unitary,clements2016optimal}.
Its action on the creation operators is
\begin{equation}
\Gamma(U)b_j^\dagger\Gamma(U)^\dagger=\sum_{i=1}^m U_{ij}b_i^\dagger.
\label{equation:optical-mode-action}
\end{equation}
Applying Eq.~\eqref{equation:optical-mode-action} to the vacuum gives Eq.~\eqref{equation:prelim-one-particle} for the single-photon states $\ket{e_j}=b_j^\dagger\ket{\mathrm{vac}}$.

\SMsection{Upper bounds for fermionic compilation}{appendix:fermionic-upper}

\SMsubsection{Single-particle encoding and unitary synthesis}{appendix:encoder}

The fermionic swap (FSWAP) and the pair flip used in the encoder are both Majorana bilinears, so by Lemma~\ref{lemma:controlled-majorana-bilinears} their controlled versions each require at most $b_W$ uses of $W$, where $b_W$ depends only on $W$.
Exchanging two modes fixes their symmetric one-particle state and negates their antisymmetric state $(\ket{10}-\ket{01})/\sqrt{2}$~\cite{bravyi2002fermionicquantumcomputation}.
The FSWAP also negates $\ket{11}$, so it equals $I-2n_-=-i\eta_1\eta_2$, where $n_-$ is the occupation operator of the antisymmetric mode and $\eta_1,\eta_2$ are its Majorana operators.

For $n\ge1$ and $D=2^n$, the encoder in Eq.~\eqref{equation:main-encoding} uses $D-1$ controlled FSWAPs and $D-1$ controlled pair flips.
Decoding applies the same self-inverse gates in reverse order.
Encoding and decoding therefore each require at most $2b_W(D-1)$ uses of $W$ and can reuse the same vacuum modes.
For a parity-preserving target $U$, the single-particle action in Eq.~\eqref{equation:prelim-one-particle} mixes only labels of the same parity, so the parity in $P$ agrees with every output label and
\begin{equation}
E_n^\dagger\bigl(I\otimes\Gamma(U)\bigr)E_n\bigl(\ket{0}_P\ket{x}\ket{e_{0^n}}\bigr)=\ket{0}_P\bigl(U\ket{x}\bigr)\ket{e_{0^n}}.
\end{equation}
The circuit thus implements both parity blocks with their relative phase, restores all ancillary modes, and requires at most $4b_W(D-1)$ uses of $W$.
For any two parity-preserving non-Gaussian gates of bounded support, the construction simulates either gate with a constant number of uses of the other.
This constant can depend on the pair.

\SMsubsection{Pure state preparation}{appendix:fermionic-state-preparation}

For a pure target of parity $\sigma$ on $n\ge2$ modes, split the data into adjacent blocks $A,B$ of $a=\lfloor n/2\rfloor$ and $b=\lceil n/2\rceil$ modes.
A singular-value decomposition within each compatible pair of local parity sectors gives
\begin{equation}
\ket{\psi}=\sum_{j=0}^{r_\psi-1}\lambda_j\ket{u_j}_A\ket{v_j}_B,\quad r_\psi\le2^a,
\label{equation:schmidt}
\end{equation}
where $r_\psi$ is the Schmidt rank and the two vectors in term $j$ have parities $\sigma_j$ and $\sigma\oplus\sigma_j$, respectively.
In the single-particle registers $R_A,R_B$ of sizes $2^a,2^b$, assign the Schmidt vectors distinct labels $a_j,b_j$ of the corresponding parities.

Order the modes as $P_A,A,P_B,B,R_A,R_B$, where $P_A,P_B$ store the local parities.
Starting from an occupation state of parity $\sigma$ and vacuum ancillas, a Gaussian unitary empties the data modes and sets the marker occupations to $(0,\sigma)$.
A Gaussian pairing rotation creates one fermion in each register, in modes $0^a$ and $0^b$, and passive mode rotations prepare $(\sum_j\lambda_j\ket{e_{a_j}}_{R_A})\ket{e_{b_0}}_{R_B}$.
For each $j$ with $\sigma_j=1$, mode $a_j$ controls the marker flip $-i\gamma_2\gamma_{2a+3}=X_{P_A}(\prod_{k\in A}Z_k)X_{P_B}$.
All modes in $A$ are empty, so these gates set the markers to $(\sigma_j,\sigma\oplus\sigma_j)$ on component $j$ without introducing a sign.
For $1\le j<r_\psi$, mode $a_j$ controls a FSWAP between $b_0$ and $b_j$.
Since $R_B$ contains one fermion, these swaps give the correlated state
\begin{equation}
\sum_{j=0}^{r_\psi-1}\lambda_j\ket{\sigma_j}_{P_A}\ket{0^a}_A\ket{\sigma\oplus\sigma_j}_{P_B}\ket{0^b}_B\ket{e_{a_j}}_{R_A}\ket{e_{b_j}}_{R_B}.
\end{equation}

Apply passive mode rotations in $R_A,R_B$ whose columns $a_j,b_j$ are the occupation-basis coefficients of $\ket{u_j},\ket{v_j}$, respectively, completing the orthonormal columns within each label parity class.
Each marker now agrees with the parity of every label in its Schmidt vector, so applying Eq.~\eqref{equation:main-encoding} in reverse on each block decodes $\ket{u_j}_A\ket{v_j}_B$.
To group the modes for decoding, move $R_A$ past $P_B,B$.
This permutation and its inverse after decoding contribute the same sign $(-1)^{\sigma\oplus\sigma_j}$ on component $j$, since $R_A$ has one fermion and $P_B,B$ have combined parity $\sigma\oplus\sigma_j$ both times.
The signs cancel, and the decoders return the markers to vacuum and the register fermions to modes $0^a,0^b$.
The inverse pairing rotation then restores both registers to vacuum, leaving the target in Eq.~\eqref{equation:schmidt}.

The two decoders require at most $2b_W(2^a+2^b-2)$ uses of $W$, and the marker flips and label swaps require at most $b_W(2r_\psi-1)$ further uses of $W$.
Thus state preparation requires $O(2^{n/2})$ uses of $W$ and $O(2^{n/2})$ ancillary modes, all returned to vacuum.
The one-mode case is Gaussian.

\SMsubsection{Exact synthesis of controlled Gaussian gates}{appendix:fermionic-controlled-proof}

For a parity-preserving unitary $G$ on the target modes, let $n_c=a_c^\dagger a_c$ be the occupation operator of a separate control mode $c$ and write $C_c(G)\coloneqq I-n_c+n_cG$.
We first establish exact reachability on the support of $W$ and one additional mode.

\begin{lemma}[Exact reachability with forward uses]
\label{lemma:exact-gate-reachability}
Let $W$ be a parity-preserving unitary on $r\ge2$ modes, embedded into $m_{\mathrm{loc}}=r+1$ modes.
Let $S$ consist of finite products of Gaussian unitaries and $W$'s, with overall phases identified.
Then $S$ equals its compact closure.
\end{lemma}

\begin{proof}
Let $d_{\mathrm{par}}=2^{m_{\mathrm{loc}}-1}$ be the dimension of each enlarged parity block.
On these two blocks, the embedded $W$ acts as direct sums of the same original parity blocks in opposite order, so the enlarged determinants agree.
Choose a scalar phase of $W$ that makes both determinants one, preserving its cost and the relative phase between parity sectors.

Let $\mc{G}_{\mathrm{F}}$ be the image of $\mathrm{Spin}(2m_{\mathrm{loc}})$ on the Fock space and $\mathfrak{g}_{\mathrm{F}}$ its Lie algebra.
For $m_{\mathrm{loc}}\ge3$, the quadratic Majorana generators are traceless in each parity block, so $\mc{G}_{\mathrm{F}}\subset\mathrm{SU}(d_{\mathrm{par}})\times\mathrm{SU}(d_{\mathrm{par}})$.
Before identifying overall phases, let $S$ be the semigroup generated by $\mc{G}_{\mathrm{F}}$ and the chosen representative of $W$, and set $G\coloneqq\overline S$.
Compactness gives positive powers of each element of $G$ approaching the identity and hence its inverse as a limit of nonnegative powers, so $G$ is a compact subgroup of $\mathrm{SU}(d_{\mathrm{par}})\times\mathrm{SU}(d_{\mathrm{par}})$.
Write $\mathfrak g\coloneqq\operatorname{Lie}(G)$ and define
\begin{equation}
\mathfrak h\coloneqq\operatorname{span}_{\R}\{\operatorname{Ad}_g X:g\in G,\ X\in\mathfrak{g}_{\mathrm{F}}\},\quad \operatorname{Ad}_g X\coloneqq gXg^{-1}.
\end{equation}
The $\operatorname{Ad}_G$-invariant subspace $\mathfrak h$ is an ideal containing $\mathfrak{g}_{\mathrm{F}}$.
With an invariant positive definite inner product on $\mathfrak g$, the orthogonal complement of $\mathfrak h$ is an ideal commuting with $\mathfrak h$.
In particular, it commutes with the quadratic generators, whose complex associative algebra is the full even Clifford algebra.
Each element of the complement is therefore scalar on each parity block, and vanishing block traces force $\mathfrak h=\mathfrak g$.

By density, $\{\operatorname{Ad}_sX:s\in S,\ X\in\mathfrak{g}_{\mathrm{F}}\}$ spans $\mathfrak g$, and each conjugate of this set by a prefix in $S$ also spans $\mathfrak g$.
Choose $s_j\in S$ and $X_j\in\mathfrak{g}_{\mathrm{F}}$ recursively so that $\operatorname{Ad}_{s_1\cdots s_j}X_j$, for $1\le j\le\ell\coloneqq\dim\mathfrak g$, form a basis.
The map
\begin{equation}
F(\theta_1,\cdots,\theta_\ell)\coloneqq s_1e^{\theta_1X_1}\cdots s_\ell e^{\theta_\ell X_\ell}
\end{equation}
takes values in $S$, since $e^{\theta_jX_j}\in \mc{G}_{\mathrm{F}}$ for every real $\theta_j$.
At the origin, the differential translated to the identity by right multiplication has the chosen basis as its columns.
The inverse function theorem gives a nonempty relatively open set $O\subset G$ contained in $S$.
For any $x\in G$, density gives $s\in S\cap xO^{-1}$, so $x=so\in S$ for some $o\in O$.
Thus $S=G$.
\end{proof}

\begin{lemma}[Controlled Majorana bilinears]
\label{lemma:controlled-majorana-bilinears}
Let $W$ be a parity-preserving unitary on $r$ modes that is non-Gaussian.
Let $Q=\pm i\eta_1\eta_2$, where $\eta_1,\eta_2$ are orthonormal Majorana directions orthogonal to both Majorana directions of the control mode $c$.
There is a constant $b_W$ depending only on $W$ such that Gaussian unitaries and at most $b_W$ uses of $W$ implement $C_c(Q)$ up to one common overall phase.
The circuit uses at most $r-1$ ancillary vacuum modes and returns them to vacuum.
\end{lemma}

\begin{proof}
Write $Z_k\coloneqq I-2n_k$ for the parity of mode $k$.
A Gaussian change of basis on the modes other than $c$ maps $Q$ to $Z_t$ or $-Z_t$ for a single mode $t$.
Since $C_c(-Z_t)=Z_c C_c(Z_t)$, it suffices to implement $C_c(Z_t)=\exp(i\pi n_c n_t)$.

Every parity-preserving unitary on one mode is Gaussian up to phase, so the hypothesis implies $r\ge2$.
Embed $W$ on the first $r$ of $m_{\mathrm{loc}}=r+1$ modes and use $\mc{G}_{\mathrm{F}},G,\mathfrak g$ from Lemma~\ref{lemma:exact-gate-reachability}.
If every $W\gamma_aW^\dagger$ for $1\le a\le2r$ were linear in the Majoranas, the anticommutation relations would give a real orthogonal action and parity preservation would make its determinant one.
The Gaussian unitary with that action would differ from $W$ only by a scalar, since the Majoranas generate the full matrix algebra.
Non-Gaussianity thus forces some $W\gamma_aW^\dagger$ to contain an odd Majorana monomial of degree at least three.

With $b=2r+1$ the first Majorana index of the spare mode, $W$ commutes with $\gamma_b$ by evenness and disjoint support, so
\begin{equation}
X\coloneqq W(\gamma_a\gamma_b)W^\dagger=(W\gamma_aW^\dagger)\gamma_b\in\mathfrak g.
\end{equation}
The inclusion uses $\gamma_a\gamma_b\in\mathfrak{g}_{\mathrm{F}}$ and invariance under conjugation by $W\in G$.
In the expansion $X=\sum_\Lambda c_\Lambda\gamma_\Lambda$ over ordered Majorana products indexed by $\Lambda\subset\{1,\cdots,2m_{\mathrm{loc}}\}$, every nonzero term contains $b$ and omits the other spare Majorana.
Choose a nonzero term of even degree $d=|\Lambda|$ with $4\le d\le2r$.

For $1\le j\le2r$, the Gaussian unitary $R_j\coloneqq\gamma_j\gamma_{2r+2}$ flips Majoranas $j$ and $2r+2$, so its conjugation negates precisely the terms containing $j$.
Starting from $Y=X$, apply the projections
\begin{equation}
Y\longmapsto\tfrac12(Y\pm R_jYR_j^\dagger),
\end{equation}
with the minus sign for $j\in\Lambda$ and the plus sign otherwise.
These $2r$ real linear projections isolate the nonzero skew Hermitian term $c_\Lambda\gamma_\Lambda$.
The projections take place in $\mathfrak g$ and serve only to establish membership in that Lie algebra.

Choose the first two modes as control and target.
Gaussian conjugation rotates the $d$ Majorana directions of the isolated monomial to the directions of the first $d/2$ modes.
After a real rescaling, this gives $i Z_1 Z_2\prod_{k=3}^{d/2} Z_k\in\mathfrak g$, with the empty product equal to $I$.
Adding the quadratic generators $-iZ_1$ and $-iZ_2$ gives
\begin{equation}
T\coloneqq i\left(-Z_1-Z_2+Z_1Z_2\prod_{k=3}^{d/2}Z_k\right)\in\mathfrak g.
\end{equation}
On the ancillary vacuum each $Z_k$ with $k\ge3$ acts as $+1$, and $I-Z_1-Z_2+Z_1 Z_2=4n_1n_2$.
Thus for every two-mode input $\ket{\psi}$,
\begin{equation}
e^{\pi T/4}\bigl(\ket{\psi}\otimes\ket{0^{r-1}}\bigr)=e^{-i\pi/4}\bigl(C_1(Z_2)\ket{\psi}\bigr)\otimes\ket{0^{r-1}}.
\end{equation}
Lemma~\ref{lemma:exact-gate-reachability} realizes $e^{\pi T/4}$ up to an overall phase with finitely many uses of $W$.
Denote this number by $b_W$.
The $r+1$ local modes consist of the control, the target, and $r-1$ vacuum modes that are returned to vacuum.
Gaussian changes of basis give every $C_c(Q)$ with the same bound, and the common phase ensures the stated action also holds on inputs entangled with other modes.
\end{proof}

\begin{proposition}
\label{proposition:controlled-fermionic-gaussian}
Let $W$ be a parity-preserving unitary on $r$ modes that is non-Gaussian.
There is a constant $b_W$ depending only on $W$ such that, for every parity-preserving Gaussian unitary $G$ on $k\ge1$ target modes, Gaussian unitaries and at most $2\lceil k/2\rceil b_W$ uses of $W$ implement $C_c(G)$ up to an overall phase.
The circuit uses at most $r$ ancillary vacuum modes and returns them to vacuum.
\end{proposition}

\begin{proof}
Decomposing the orthogonal Majorana action in Eq.~\eqref{equation:prelim-fermionic-majorana-action} into rotation planes gives $G=e^{i\phi}\prod_{j=1}^k e^{\theta_j u_jv_j}$ for real angles $\phi,\theta_j$ and an orthonormal basis of target Majorana operators $u_1,v_1,\cdots,u_k,v_k$.
The Gaussian gate $e^{i\phi n_c}$ supplies the specified phase of $G$ on the occupied control branch.
Group the rotation planes in pairs and consider one pair with generator $A=\theta_1u_1v_1+\theta_2u_2v_2$.
The involution $Q=-iu_1u_2$ negates one Majorana in each plane under conjugation, so $QAQ=-A$.
With $F=e^{A/2}$, this gives
\begin{equation}
C_c(e^A)=F C_c(Q) F^\dagger C_c(Q),
\end{equation}
since the right-hand side acts as $I$ when $c$ is empty and as $FQF^\dagger Q=F^2=e^A$ when it is occupied.
For an unpaired plane, use $A=\theta_j u_jv_j$ and $Q=-iu_j\xi$, taking $\xi$ from another target plane or, when $k=1$, from one additional vacuum mode.
The same identity holds and acts trivially on the mode supplying $\xi$ after the two controlled gates.
Choose $b_W$ as in Lemma~\ref{lemma:controlled-majorana-bilinears}.
The two controlled gates in each of the $\lceil k/2\rceil$ groups require at most $2b_W$ uses of $W$ and can reuse the same $r-1$ vacuum modes, with one extra vacuum mode for $\xi$ when $k=1$.
\end{proof}

\SMsection{Lower bounds for fermionic compilation}{appendix:fermionic-lower}

\begin{theorem}[Fermionic lower bounds]
\label{theorem:fermionic-lower}
Fix a parity-preserving unitary $W$ on a bounded number of modes and allow Gaussian ancillas on any finite number of modes, adaptive Gaussian processing, and postselection.
With $D=2^n$, for sufficiently large $D$ the number $t$ of uses of $W$ for exact synthesis on $n$ modes satisfies
\begin{equation}
t\ge c\sqrt D\quad\text{for states},\quad t\ge cD\quad\text{for unitaries},
\end{equation}
for almost every pure state in either parity sector and almost every parity-preserving unitary, respectively.
These bounds hold on every successful branch of positive probability and, for continuous measurement records, almost surely conditional on success.
The positive constant $c$ and size threshold depend only on the support size of $W$.
\end{theorem}

\SMsubsection{Gaussian branches and parameter bounds}{appendix:kernel}

We first put a successful branch into the form shown in Fig.~\ref{figure:lower-bound}. Purify any mixed Gaussian ancillary states and refine discarded modes, including purifying modes, by Gaussian measurements. Condition on the refined measurement record, which determines all adaptive choices. Fresh Gaussian ancillas implement resets and allow measured modes and dilation pointers to remain idle until their final projections. This gives a sequence of Gaussian unitaries and uses of $W$, followed by Gaussian outcome contractions. The refinement preserves the number of uses of $W$.

In the fermionic model, discarded modes can be refined by occupation measurements. To defer an occupation projector $\ketbra{b}{b}$, swap the measured mode with a fresh mode prepared in $\ket{b}$ and postselect the old state in the fresh mode onto $\ket{b}$ at the end. On this outcome the fermionic swap contributes only the sign $(-1)^b$. Absorb pure Gaussian ancillary preparations into the Gaussian layers. An odd pure Gaussian ancillary state can be paired with an occupied spectator so that the joint pure Gaussian input is even and can be prepared from vacuum. Postselect that spectator onto its occupied state at the end.

\begin{lemma}[Support of the conjugated gates]
\label{lemma:ferm-core-compression}
Consider a fermionic circuit on $m_{\mathrm{tot}}$ modes with $m$ data modes, vacuum ancillas, and $t$ uses of a parity-preserving gate $W$ of bounded support. After moving all Gaussian unitaries to the end and making a passive basis change that fixes the data modes, the conjugated gates act on a common register of $m_{\mathrm{eff}}$ modes with
\begin{equation}
m_{\mathrm{eff}}\le m_{\mathrm{tot}},\qquad m_{\mathrm{eff}}\le m+O(t).
\label{equation:effective-modes}
\end{equation}
The complementary modes stay in vacuum until the final Gaussian unitary. Each conjugated gate depends polynomially on $O(m_{\mathrm{eff}})$ real coefficients.
\end{lemma}

\begin{proof}
Write the circuit as Gaussian layers $G_0,\cdots,G_t$ interleaved with uses of $W$. With $G_{<j}=G_{j-1}\cdots G_0$ and $G_{\mathrm{fin}}=G_t\cdots G_0$, the total unitary is $G_{\mathrm{fin}}W_t\cdots W_1$, where $W_j=G_{<j}^\dagger W G_{<j}$ with the support of the $j$th use of $W$ understood. Expand $W$ in even monomials of the Majoranas on its support. Conjugation replaces each of those Majoranas with a real linear combination of the full register's Majoranas.

A Hermitian linear combination of Majoranas can be written as $a^\dagger(v)+a(v)$ for a single complex mode direction $v$. Each use of $W$ contributes a bounded number of such directions, so their span together with the data input modes obeys \eqref{equation:effective-modes}. Choose an orthonormal basis of this span beginning with the data modes, and extend it to the full register. This passive basis change preserves the vacuum. Each even Majorana monomial on the effective modes acts trivially on the complement. Each transformed Majorana has $2m_{\mathrm{eff}}$ real coefficients in the new basis, and substitution into the finite monomial expansion gives the claimed polynomial dependence.
\end{proof}

\begin{lemma}[Fermionic Gaussian kernels]
\label{lemma:ferm-gaussian-kernels}
A nonzero conditional fermionic Gaussian map $K$ from $m_{\mathrm{eff}}$ input to $m$ output modes, obtained by adjoining vacuum modes, applying a Gaussian unitary, and projecting the unretained modes onto an occupation pattern, is described up to a nonzero scalar by finitely many polynomial charts with at most $(m_{\mathrm{eff}}+m)(m_{\mathrm{eff}}+m-1)$ real parameters each.
\end{lemma}

\begin{proof}
Write $K_{xy}=\bra{x,z}G\ket{y,0}$, where $z$ is the occupation pattern on the unretained modes. Order the retained modes first in the output basis and the effective modes first in the input basis. Choose an entry $K_{x_0y_0}\ne0$ and set $\ket{R}=\ket{y_0,0}$, $\ket{L}=G^\dagger\ket{x_0,z}$, and $\omega(X)=\bra{L}X\ket{R}/\braket{L}{R}$. The states $\ket{L}$ and $\ket{R}$ are pure Gaussian and have the same parity. For each input mode $i$, take $\gamma_i^{\rm in}=a_i+a_i^\dagger$.
The product of these Majoranas over the sites where $y$ and $y_0$ differ, ordered by increasing mode index, takes $\ket{y_0,0}$ to $\ket{y,0}$ up to a sign. The same construction on the output gives $\bra{x,z}$ from $\bra{x_0,z}$. Moving the output Majoranas through $G$ thus expresses each ratio $K_{xy}/K_{x_0y_0}$, up to a sign determined by the occupation strings and the ordering, as $\omega$ of an ordered subproduct of the list
\begin{equation}
G^\dagger\gamma_1^{\rm out}G,\cdots,G^\dagger\gamma_m^{\rm out}G,\ 
\gamma_1^{\rm in},\cdots,\gamma_{m_{\mathrm{eff}}}^{\rm in}.
\end{equation}
The ordering of the output flips is chosen consistently with the order of bra multiplication.

For completeness, Wick's rule for this transition functional follows from the annihilator spaces of $\ket R$ and $\bra L$ in the complex span of the Majoranas. Both have half the dimension of that span. They intersect trivially because a nonzero common annihilator $c$ would give $\bra L\{c,c^\dagger\}\ket R=0$, whereas $\{c,c^\dagger\}$ is a positive scalar times $I$ and $\braket{L}{R}\ne0$. Decompose the first operator in any product along the two spaces. Its left-annihilating component drops out, and anticommuting its right-annihilating component through the remaining operators gives the recursion in terms of two-point contractions. Odd moments vanish by parity. Thus every even moment is a Pfaffian of the pairwise contractions, with the empty moment equal to one. The list has $m_{\mathrm{eff}}+m$ entries, so the at most $\binom{m_{\mathrm{eff}}+m}{2}$ complex contractions suffice. Allowing them to range freely yields polynomial parameterizations containing the normalized matrices. The finitely many choices of the reference entry cover all nonzero maps.
\end{proof}

For a branch with $t$ uses of $W$, Lemma~\ref{lemma:ferm-core-compression} gives the unnormalized output on input $\ket{y}$ as
\begin{equation}
K W_t\cdots W_1(\ket{y}\otimes\ket{\mathrm{vac}}),
\label{equation:conditional-branch-normal-form}
\end{equation}
where the vacuum factor fills the extra effective modes, and $K$ is obtained by contracting the final Gaussian unitary with the complementary input vacuum and the deferred output projections. Lemma~\ref{lemma:ferm-gaussian-kernels} bounds the real parameter dimension for these matrix elements, up to scalar, by
\begin{equation}
O\bigl(tm_{\mathrm{eff}}+(m_{\mathrm{eff}}+m)^2\bigr)=O((m+t)^2).
\label{equation:kernel-dim}
\end{equation}
The state-preparation column and the branch matrix both depend smoothly on the parameters in projective charts of their nonzero rays. This count is independent of ancillary width and the length of the Gaussian part of the circuit.

\SMsubsection{Dimension comparison and adaptive protocols}{appendix:fermionic-dimension-protocols}

\begin{proof}[Proof of Theorem~\ref{theorem:fermionic-lower}]
On a refined exact preparation branch the nonzero output has the target ray, by positivity of the contributions summing or integrating to the pure output. For exact unitary synthesis let $L_b$ be the input--output operator of refinement $b$. For each pure input $\ket{\psi}$, the same argument gives $L_b\ket{\psi}\in\Span(U\ket{\psi})$ for almost every successful refinement. Apply this to a basis and all its pairwise sums. The finite intersection of these conditions and linearity imply $L_b=c_bU$ almost everywhere. When a pair spans both fermionic parity sectors, the two basis states can be paired with opposite occupations of an idle reference mode so that the superposition has definite total parity. The outcome probability or density $|c_b|^2$ is independent of the input.

The image of a smooth parameterization from a space of dimension $p$ has Hausdorff dimension at most $p$, by covering the coordinate domain by countably many compact patches on which the map is Lipschitz. Such an image meets any target manifold of larger dimension in a set of zero induced measure. Here the target dimensions are $D-2$ for pure state rays in either parity sector and $D^2/2-1$ for parity-preserving unitaries modulo a common global phase. For each $t$, the possible effective dimensions and the projective charts give a countable cover. Comparing Eq.~\eqref{equation:kernel-dim} with the two target dimensions gives $\Omega(\sqrt D)$ uses of $W$ for states and $\Omega(D)$ uses of $W$ for unitaries. For each $W$, taking the countable union of the exceptional sets over all possible numbers of uses of $W$ gives a common full-measure set of targets across protocols and outcomes.
\end{proof}

For an almost surely terminating exact protocol, averaging the branch counts gives the same lower bound on the expected number of uses of $W$, conditioned on success when postselection is used.

For a circuit with $t$ uses of $W$, running it with probability $p$ and otherwise declaring immediate failure gives an unconditional expected count of $pt$, which can be arbitrarily small.

\SMsection{Upper bounds for bosonic compilation}{appendix:bosonic-upper}

\SMsubsection{Single-particle encoding and unitary synthesis}{appendix:bos-encoder}

The encoder in Eq.~\eqref{equation:main-boson-encoder} implements photonic state joining~\cite{passaro2013joiningsplittingquantumstatesphotons} using $2q$ data modes and $2^q$ modes in the single-particle register.
At the input to each controlled swap, the control mode contains at most one photon and the two target modes together contain at most one photon.
By Corollary~\ref{corollary:bos-controlled-swap}, passive optics and at most $b_W$ uses of $W$ implement each such swap by borrowing $\ell-2$ other dual-rail qubits with one photon each, where $\ell\ge2$ and $b_W$ depend only on $W$.
Each encoder or decoder uses two controlled swaps per prefix at steps $j=2,\cdots,q$, for a total of
\begin{equation}
t_q\coloneqq 2\sum_{j=2}^q 2^{j-1}=2(2^q-2).
\label{equation:bos-encoder-count}
\end{equation}
For $q=1$, the encoder is a passive mode permutation and any logical unitary is a two-mode interferometer.

At each step $j=2,\cdots,q$ of the encoder, the $j-2$ reset qubits and the $q-j$ unprocessed qubits provide $q-2$ occupied rail pairs outside each controlled swap.
For $q\ge\ell$, these rail pairs provide the required $\ell-2$ photons at every step of the encoder and its reverse.
For any logical unitary $U$, the circuit $E_q^\dagger\Gamma(U)E_q$ implements $U$ exactly up to an overall phase with at most $2b_W t_q=4b_W(2^q-2)$ uses of $W$ by Eq.~\eqref{equation:bos-encoder-count}.
Reusing the vacuum ancillas for each swap gives a total of $2^q+2q+O(1)$ modes.

\SMsubsection{Pure state preparation}{appendix:bosonic-state-preparation}

For $q\ge2\ell$, we apply the Schmidt construction of Sec.~\ref{section:settings-and-main-results} to blocks of $a\coloneqq\lceil q/2\rceil$ and $b\coloneqq\lfloor q/2\rfloor$ qubits.
A passive permutation of the dual-rail zero state moves the photon from the first qubit of each block to mode $0$ of its single-particle register, leaving the other $q-2$ qubits in $\ket{0}_{\rm L}$.
For a target of Schmidt rank $r_\psi\le2^b$, each of the $r_\psi-1$ controlled swaps that correlate the Schmidt labels borrows $\ell-2$ of these photons.
Since each block has at least $\ell$ qubits, its decoder can borrow the required photons within that block as in Sec.~\ref{appendix:bos-encoder}.
Using the decoder counts from Eq.~\eqref{equation:bos-encoder-count}, the circuit requires at most $b_W(r_\psi-1+t_a+t_b)=O(2^{q/2})$ uses of $W$ and $2^a+2^b+2q+O(1)=O(2^{q/2}+q)$ modes in total.
Both constructions use exactly the $q$ input photons and restore all ancillary modes to vacuum.

\SMsubsection{Exact synthesis of controlled swaps}{appendix:bos-fixed-gates}

\begin{lemma}[Exact synthesis in a photon-number sector]
\label{lemma:bos-fixed-sector}
Let $W$ be a non-Gaussian unitary on $r$ modes that preserves total photon number.
There is a photon number $\ell\ge2$ such that, for every $m_{\mathrm{loc}}\ge\max\{r+1,3\}$, passive optics and finitely many uses of $W$ can implement any isometry from a proper subspace of the $\ell$-photon sector on $m_{\mathrm{loc}}$ modes into that sector.
The circuit implements the isometry and the identity on every lower photon sector exactly up to one common overall phase.
\end{lemma}

\begin{proof}
Let $W^{(j)}$ denote the $j$-photon block of $W$, with vacuum phase $w_0=W^{(0)}$.
The normalized gate
\begin{equation}
\widetilde W\coloneqq w_0^{-1}\Gamma(w_0(W^{(1)})^\dagger)W
\end{equation}
fixes the vacuum and the one-photon sector and can be implemented up to a global phase by applying $W$ once followed by $\Gamma(w_0(W^{(1)})^\dagger)$.
If all higher photon blocks were also identities, $W$ would equal a mode rotation up to phase and hence would be Gaussian.
Let $\ell\ge2$ be the first photon sector with $\widetilde W^{(\ell)}\ne I$, so $\widetilde W$ is the identity on every lower sector.

Embed $\widetilde W$ into $m_{\mathrm{loc}}$ modes and select a mode $z$ outside its original support.
On the $\ell$-photon sector the embedded gate is nontrivial, but fixes the vector with all photons in $z$ and every symmetrized vector with $\ell-1$ photons in $z$ and one in an arbitrary one-particle vector $v$.
If this restriction agreed up to phase with the $\ell$-photon action $\Gamma_\ell(U)$ of a mode rotation, we could absorb the phase into the one-particle unitary $U$.
Fixing the first vector forces $Uz=\lambda z$ with $\lambda^\ell=1$.
Fixing the vectors with $v\perp z$ then gives $Uv=\lambda v$.
Consequently $U=\lambda I$ and $\Gamma_\ell(U)=I$, a contradiction.
By the bosonic universality theorem for more than two modes~\cite{oszmaniec2017universalextensionsrestrictedclassesquantumoperations}, the transformations generated on this sector are dense in $\mathrm{U}(d_\ell)$, where $d_\ell\coloneqq\binom{m_{\mathrm{loc}}+\ell-1}{\ell}$ is its dimension.

To obtain an exact circuit while controlling the lower sectors, track the one-photon and $\ell$-photon actions jointly.
Let $H\coloneqq\{(U,\Gamma_\ell(U)):U\in\mathrm{U}(m_{\mathrm{loc}})\}$ be the joint action of the passive subgroup, and let $S$ be the semigroup generated by $H$ and $(I,\widetilde W^{(\ell)})$.
Its closure $G\subset\mathrm{U}(m_{\mathrm{loc}})\times\mathrm{U}(d_\ell)$ is a compact group, because compact recurrence supplies inverses as limits of positive powers.
Its Lie algebra $\mathfrak g$ projects onto both $\mathfrak u(m_{\mathrm{loc}})$ and $\mathfrak u(d_\ell)$.
The kernel of the first projection, viewed in the second factor, is an ideal $\mathfrak n$ of $\mathfrak u(d_\ell)$.
Surjectivity of the two projections gives
\begin{equation}
\dim\mathfrak n=\dim\mathfrak g-m_{\mathrm{loc}}^2\ge d_\ell^2-m_{\mathrm{loc}}^2>1,
\end{equation}
where the last inequality holds for $m_{\mathrm{loc}}\ge3$ and $\ell\ge2$.
The center of $\mathfrak u(d_\ell)$ is one dimensional and $\mathfrak{su}(d_\ell)$ is simple, so $\mathfrak n$ contains $\mathfrak{su}(d_\ell)$.
Thus
\begin{equation}
\{0\}\oplus\mathfrak{su}(d_\ell)\subset\mathfrak g.
\end{equation}
An isometry on a proper subspace of dimension $p<d_\ell$ is specified by the images of an orthonormal basis of that subspace.
To track these images together with the one-photon action, consider
\begin{equation}
X\coloneqq\mathrm{U}(m_{\mathrm{loc}})\times\operatorname{St}_p(\mbb{C}^{d_\ell}),
\end{equation}
where $\operatorname{St}_p$ denotes the manifold of ordered orthonormal $p$-frames.
The subgroup $H$ sets the first coordinate, and the independent $\mathrm{SU}(d_\ell)$ action sets the second since $p<d_\ell$ leaves a column to adjust the determinant.
Hence $G$ acts transitively on $X$.

Let $\mathfrak a$ be the real span of $\operatorname{Ad}_gY$ for $g\in G$ and $Y\in\operatorname{Lie}(H)$.
It is an ideal of $\mathfrak g$ by $\operatorname{Ad}_G$ invariance and contains $\operatorname{Lie}(H)$.
Commuting a generator in $\operatorname{Lie}(H)$ whose $\ell$-photon block is noncentral with a suitable generator of $\{0\}\oplus\mathfrak{su}(d_\ell)$ gives a nonzero element in the intersection of the two ideals.
By simplicity, $\mathfrak a$ therefore contains $\{0\}\oplus\mathfrak{su}(d_\ell)$.
Together with the directions generated by $H$, these generators span the tangent space to $X$ at every point.

Let $x_0\in X$ consist of the identity one-photon action and the chosen input basis.
By density, the conjugates $\operatorname{Ad}_sY$ with $s\in S$ and $Y\in\operatorname{Lie}(H)$ also span $\mathfrak a$.
The product construction in the proof of Lemma~\ref{lemma:exact-gate-reachability} therefore gives a family of forward circuits whose action on $x_0$ has surjective differential with respect to the passive gate parameters at zero.
Its image contains a nonempty open set $O\subset X$ of exactly reachable points.
For each $x\in X$, transitivity and density give an $s\in S$ with $s^{-1}x\in O$, so applying $s$ to this reachable point reaches $x$ exactly.
In particular, we can realize the requested output frame with identity one-photon action.
Since $\widetilde W$ fixes every lower sector, the net passive transformation on those sectors is determined by the one-photon action and is therefore the identity.
Restoring the vacuum phase of $W$ changes all sectors by the same overall phase.
\end{proof}

\begin{corollary}
\label{corollary:bos-controlled-swap}
Let $W$ and $\ell$ be as in Lemma~\ref{lemma:bos-fixed-sector}, and suppose $\ell-2$ additional dual-rail qubits are available, each with one photon.
On the subspace $n_c\le1$ and $n_u+n_v\le1$, passive optics and at most $b_W$ uses of $W$ implement the swap of modes $u,v$ controlled by the occupation of mode $c$, where $b_W$ depends only on $W$.
The implementation is exact up to one common overall phase, acts as the identity on the borrowed qubits, and restores all vacuum ancillas.
\end{corollary}

\begin{proof}
The only allowed two-photon states of $c,u,v$ in the occupation basis are $\ket{110}$ and $\ket{101}$.
Including both rails of each borrowed qubit, we prescribe the exchanges
\begin{equation}
\ket{110}\ket{y}_{\mathrm L}\longleftrightarrow\ket{101}\ket{y}_{\mathrm L},\quad y\in\{0,1\}^{\ell-2}.
\end{equation}
With enough vacuum modes, these $2^{\ell-1}$ basis vectors span a proper subspace of the $\ell$-photon sector and the mode count meets the hypothesis of Lemma~\ref{lemma:bos-fixed-sector}.
The lemma implements the prescribed exchanges exactly and fixes every lower sector with the same overall phase, covering all inputs with at most one photon in $c,u,v$ combined.
Since the exchange preserves every $\ket{y}_{\mathrm L}$ with the same phase, it acts as the identity on the borrowed qubits even when they are entangled with other modes.
The prescribed map and the number of modes depend only on $W$, giving the bound $b_W$.
\end{proof}

For any $\ell\ge2$, the single-mode gate $I-2\ketbra{\ell}{\ell}$ acts as the identity whenever fewer than $\ell$ photons are available, so the photon threshold can depend on $W$.

\SMsection{Lower bounds for bosonic compilation}{appendix:bos-lower}

\begin{theorem}[Bosonic lower bounds]
\label{theorem:bos-code-lower}
Fix a photon-number-preserving unitary $W$ on a bounded number of modes and allow Gaussian ancillas on any finite number of modes, adaptive Gaussian processing, and postselection.
With $D=2^q$, for sufficiently large $D$ the number $t$ of uses of $W$ for exact synthesis on $q$ dual-rail qubits satisfies
\begin{equation}
t\ge c\sqrt D\quad\text{for states},\quad t\ge cD\quad\text{for unitaries},
\end{equation}
for almost every pure state and almost every unitary, respectively.
These bounds hold on every successful branch of positive probability and, for continuous measurement records, almost surely conditional on success.
The positive constant $c$ and size threshold depend only on the support size of $W$.
\end{theorem}

We use the same branch form as in Eq.~\eqref{equation:conditional-branch-normal-form}. The bosonic argument uses affine canonical mode operators for the conjugated gates and a distributional kernel for the conditional Gaussian map.

\SMsubsection{Gaussian branches and parameter bounds}{appendix:bos-kerr-exact-lower}

The lower bound imposes no restriction on intermediate photon number.
We consider $q$ dual-rail qubits on $m=2q$ data modes, with logical basis $\ket{x}_{\rm L}$ for $x\in\{0,1\}^q$.
We specify the input and retained output mode labels in advance.
State preparation starts from $\ket{0}_{\rm L}^{\otimes q}$.

\begin{lemma}
\label{lemma:bos-core-compression}
A unitary circuit on $m_{\mathrm{tot}}$ modes, with arbitrary Gaussian unitaries, vacuum helpers, and $t$ uses of a unitary $W$ of bounded support preserving photon number, admits a Gaussian normal form in which all $t$ conjugated gates act on a common mode subspace containing the $m$ input modes and of dimension
\begin{equation}
m_{\mathrm{eff}}\le m_{\mathrm{tot}},\qquad m_{\mathrm{eff}}\le m+O(t).
\end{equation}
Each conjugated gate is implementable on these $m_{\mathrm{eff}}$ modes with one use of $W$ and Gaussian controls.
The remaining modes stay in vacuum until the final Gaussian unitary.
\end{lemma}

\begin{proof}
As in Lemma~\ref{lemma:ferm-core-compression}, put all Gaussian unitaries at the end. By Eq.~\eqref{equation:optical-gaussian-action}, the $j$th conjugated gate is supported on the affine modes
\begin{equation}
c_{j\alpha}=u_{j\alpha}\boldsymbol b+v_{j\alpha}\boldsymbol b^\dagger+\beta_{j\alpha},
\end{equation}
where $\boldsymbol b$ and $\boldsymbol b^\dagger$ are columns of annihilation and creation operators, $u_{j\alpha},v_{j\alpha}\in\mbb{C}^{1\times m_{\mathrm{tot}}}$, and $\beta_{j\alpha}\in\mbb{C}$.
For each $j$, the relations $[c_{j\alpha},c_{j\gamma}^\dagger]=\delta_{\alpha\gamma}I$ and $[c_{j\alpha},c_{j\gamma}]=0$ make the affine modes on the support of $W$ a canonical frame.
Conjugating $W$ by a Gaussian unitary that carries its support annihilators to this frame defines $W(\boldsymbol c_j)$.
For two completions $U_1,U_2$, the unitary $U_2^\dagger U_1$ fixes the support Weyl algebra and acts only on complementary modes, so it commutes with $W\otimes I$.

In the same passive basis construction, include the input modes and the two mode directions $u_{j\alpha}^\dagger$ and $v_{j\alpha}^T$ for each $c_{j\alpha}$. The displacement is a scalar. Their span defines the effective mode subspace and obeys \eqref{equation:effective-modes}, since each use of $W$ contributes only a bounded number of directions. Its complement stays in vacuum until the final Gaussian unitary.

The real quadrature rows of each frame span a nondegenerate symplectic subspace.
A symplectic basis of its symplectic orthogonal complement and an extension of the displacement give a finite Gaussian completion on the effective mode subspace.
Each $W(\boldsymbol c_j)$ therefore requires one use of $W$ and free Gaussian conjugation on $m_{\mathrm{eff}}$ modes.
\end{proof}

For a register of $m'$ modes, let $N_{m'}\coloneqq\sum_{j=1}^{m'} b_j^\dagger b_j$ and identify $\mc{S}_{m'}\coloneqq\bigcap_{\ell=0}^{\infty}\operatorname{Dom}(1+N_{m'})^\ell$ with the Schwartz space on $\mbb{R}^{m'}$, equipped with seminorms
\begin{equation}
\|\psi\|_\ell\coloneqq\|(1+N_{m'})^\ell\psi\|,\quad \ell=0,1,\cdots.
\end{equation}
Every finite Fock vector belongs to this space.
We write $\mc{S}'_{m'}$ for its space of tempered distributions.

\begin{lemma}
\label{lemma:bos-gaussian-kernels}
Let $K$ be a conditional Gaussian map from $m_{\mathrm{eff}}$ input modes to $m$ retained modes obtained from a finite Gaussian circuit with normalizable pure Gaussian ancillary inputs and Gaussian outcome bras, including destructive homodyne outcomes.
Its distribution kernel belongs to a family covered by countably many smooth parameterizations in $\mc{S}'_{m_{\mathrm{eff}}+m}$, each with $O((m_{\mathrm{eff}}+m)^2)$ real parameters.
The parameter bound is independent of the number of ancillary modes and measurements in the circuit.
\end{lemma}

\begin{proof}
Set $N=m_{\mathrm{eff}}+m$ and fix a support dimension $0\le j\le N$.
Choose a real matrix $V\in\mbb{R}^{N\times j}$ of rank $j$, vectors $x_0\in\mbb{R}^N$ and $a,v\in\mbb{R}^j$, a real matrix $B\in\mbb{R}^{j\times j}$, a real symmetric matrix $H\in\mbb{R}^{j\times j}$, and a scalar $c\in\mbb{C}$.
Define a distribution $\mathcal K$ by
\begin{equation}
\mathcal K[f]=c\int_{\mbb{R}^j}e^{-\|B(u-a)\|^2/2+iu^THu/2+iv^Tu}f(Vu+x_0)\,du
\end{equation}
for $f\in\mc{S}_N$, with $\mathcal K[f]=cf(x_0)$ when $j=0$.
The distribution is supported on $x_0+\operatorname{ran}V$ and uses at most $(5N^2+7N+4)/2$ real parameters.

Gaussian integration places every Gaussian circuit kernel in this family.
The identity kernel is a delta distribution on the diagonal, normalizable Gaussian inputs and outcome bras have Gaussian wave functions, and homodyne bras evaluate a position coordinate after a Gaussian rotation.
The generators of Gaussian unitaries preserve this form and consist of translations, modulations, invertible real coordinate changes, quadratic phase multiplication, and partial Fourier transforms.

For a degenerate Fourier integral, write the quadratic matrix in the integrated coordinates as $P-iQ$, with $P,Q$ real symmetric and $P\ge0$.
Its nullspace is
\begin{equation}
\ker_{\mbb{C}}(P-iQ)=\bigl(\ker_{\mbb{R}}P\cap\ker_{\mbb{R}}Q\bigr)\otimes_{\mbb{R}}\mbb{C}.
\end{equation}
For $(P-iQ)z=0$, taking the real part of $z^\dagger(P-iQ)z$ gives $z^\dagger Pz=0$, hence $Pz=Qz=0$.
Positive semidefiniteness of the full damping form makes its cross terms with the real null directions vanish.
Integration along these directions gives ordinary real affine delta constraints, and integration on their complement gives a Gaussian Schur complement and a scalar factor.
No polynomial prefactors or derivatives of delta distributions are introduced.

Gaussian unitaries, tensoring with normalizable Gaussian vectors, Gaussian bra contractions, and homodyne evaluations act continuously between Schwartz spaces.
Thus the resulting kernel is tempered, and any Gaussian exponential on its affine support has nonnegative real damping.
Its real linear coefficient lies in the range of that damping matrix, since otherwise it grows exponentially in an undamped direction.
Completing the square gives the center $a$, and the positive semidefinite damping matrix factors as $B^TB$.

Before a homodyne contraction, the affine kernel support of a nonzero map projects onto the full output coordinate space, since a Schwartz output supported on a proper affine subspace vanishes.
Evaluation at the measured coordinates is therefore transverse and preserves the displayed kernel form.

On compact parameter sets with $V$ of full rank, $\|Vu+x_0\|$ grows at least linearly in $\|u\|$ as $\|u\|\to\infty$, uniformly in the parameters.
The exponential has modulus at most one, and each parameter derivative adds only polynomial factors and derivatives of the test function.
Schwartz decay bounds these derivatives locally uniformly, including for singular $B$ and test functions varying smoothly in Schwartz space, so the distribution family is smooth.
The finitely many support dimensions $j$ and countably many charts of the full-rank domains give the asserted cover.
\end{proof}

\begin{lemma}
\label{lemma:bos-adaptive-kernel}
Fix $q$ dual-rail qubits and a photon-number-preserving unitary $W$ on a bounded number of modes.
State preparation starts from $\ket{0}_{\rm L}^{\otimes q}$.
Allow arbitrary finite Gaussian ancillary systems, adaptive Gaussian measurements and controls, Gaussian resets, discarding, and postselection.
For branches with $t$ uses of $W$, the nonzero projected state vectors and branch matrices in the logical basis have rays in countably many smooth images of real dimension at most $O((m+t)^2)$ with $m=2q$.
\end{lemma}

\begin{proof}
Condition on the measurement record as in Sec.~\ref{appendix:kernel}. Bosonic Gaussian instruments admit pure Gaussian outcome refinements, including homodyne and heterodyne measurements~\cite{giedke2002characterizationgaussianoperationsdistillation}. Refine discarded and purifying modes by heterodyne measurements and absorb each pure Gaussian ancillary preparation from vacuum into the Gaussian layers. Lemma~\ref{lemma:bos-core-compression} puts the resulting branch into Eq.~\eqref{equation:conditional-branch-normal-form} with $m=2q$, $m_{\mathrm{eff}}\le m+O(t)$, and $W_j=W(\boldsymbol c_j)$ on the logical inputs $\ket{y}_{\rm L}$.

Each affine canonical frame on $m_{\mathrm{eff}}$ modes has $O(m_{\mathrm{eff}})$ real parameters and admits smooth local Gaussian completions.
Since $W$ preserves photon number, its extension to $m_{\mathrm{eff}}$ modes commutes with $N_{m_{\mathrm{eff}}}$ and satisfies
\begin{equation}
\|W\psi\|_\ell=\|W^\dagger\psi\|_\ell=\|\psi\|_\ell.
\end{equation}
The affine quadrature action of a Gaussian unitary $G$ gives $\|G\psi\|_\ell\le C_\ell(G)\|\psi\|_\ell$ on $\mc{S}_{m_{\mathrm{eff}}}$, with locally bounded constants in the Gaussian parameters.
Differentiation inserts linear or quadratic operators, which are continuous between suitable Schwartz seminorms.
Iteration gives a smooth Gaussian action on $\mc{S}_{m_{\mathrm{eff}}}$ and hence smooth dependence of each conjugated gate and finite word, using only number preservation for $W$.

For frame parameters $\theta$, let $\psi_y(\theta)\in\mc{S}_{m_{\mathrm{eff}}}$ be the result of applying $W_t\cdots W_1$ to $\ket{y}_{\rm L}\otimes\ket{\mathrm{vac}}$, and let $\phi_x\in\mc{S}_m$ be the wave function of $\ket{x}_{\rm L}$.
The logical matrix element is the distribution pairing
\begin{equation}
F_{xy}(\theta,\eta)=\mathcal{K}_\eta[\overline{\phi_x}\otimes\psi_y(\theta)],
\end{equation}
where $\eta$ parameterizes the distribution kernel $\mathcal{K}_\eta\in\mc{S}'_{m+m_{\mathrm{eff}}}$ of $K$.
By Lemma~\ref{lemma:bos-gaussian-kernels} and the smooth dependence of $\psi_y(\theta)$ in Schwartz space, the pairing is smooth in $(\theta,\eta)$.
State preparation uses the column $y=0^q$, and taking the ray of a nonzero amplitude vector or vectorized branch matrix is smooth on each projective chart.
The total number of parameters is bounded by
\begin{equation}
O\bigl(tm_{\mathrm{eff}}+(m_{\mathrm{eff}}+m)^2\bigr)=O((m+t)^2).
\label{equation:bos-gaussian-dimension}
\end{equation}
The frame and kernel parameters account for all measurement records and feedforward rules regardless of ancillary width, free operation count, or record length.
For each $t$, there are finitely many effective dimensions and countably many frame, kernel, and projective charts, giving the claimed cover.
\end{proof}

\SMsubsection{Dimension comparison and adaptive protocols}{appendix:bos-dimension-protocols}

\begin{proof}[Proof of Theorem~\ref{theorem:bos-code-lower}]
The branchwise rank-one and measure-zero arguments of Sec.~\ref{appendix:fermionic-dimension-protocols} apply on the dual-rail input space. For continuous records, postselection accepts an event of positive outcome probability, with individual homodyne values labeling Kraus densities. The logical state ray and unitary manifolds have real dimensions $2D-2$ and $D^2-1$, respectively. Comparing with Eq.~\eqref{equation:bos-gaussian-dimension} gives $\Omega(\sqrt D)$ uses of $W$ for states and $\Omega(D)$ uses of $W$ for unitaries.
For each $W$, taking the countable union of exceptional sets over all possible numbers of uses of $W$ gives a common full-measure set of targets.
\end{proof}

For an almost surely terminating exact protocol, averaging the branch counts gives the same lower bound on the expected number of uses of $W$, conditioned on success when postselection is used.

\SMsection{Mode--gate tradeoff}{appendix:mode-gate-tradeoff}

\begin{proof}[Proof of Theorem~\ref{theorem:mode-gate-tradeoff}]
Fix the total mode count $m_{\mathrm{tot}}$ and the number $t$ of uses of $W$ in a coherent circuit.
A Gaussian change of basis brings the ancillary input to a product of single-mode states, and each mixed factor has a two-mode Gaussian purification.
Thus we can purify the ancillary input with at most one extra mode per ancillary mode, giving an enlarged width $M\le2m_{\mathrm{tot}}$.
The purifying modes remain idle during the original circuit, so the enlarged circuit still uses $W$ exactly $t$ times.
Absorb preparation of the pure Gaussian ancillary input into the first Gaussian layer, using an occupation state of the appropriate parity as a reference for fermions and vacuum for bosons.

Contract the final ancillary and purifying modes with a fixed bra from their occupation basis.
By the purity and rank-one arguments of Sec.~\ref{appendix:fermionic-dimension-protocols}, every nonzero contraction is proportional to the target state vector or to the target unitary on the allowed input space.
At least one contraction is nonzero by completeness of the occupation basis.
For fixed $m_{\mathrm{tot}}$, the enlarged width and the input and output occupation patterns give only countably many choices.

Move all Gaussian unitaries to the end, as in the lower-bound proofs above.
Each conjugated gate $W_j$ is specified by the transformed mode operators on the support of $W$ and uses $O(M)$ real parameters, and the final Gaussian unitary contributes $O(M^2)$ parameters.
For fixed occupation patterns, the contracted amplitudes depend smoothly on these parameters in local charts.
In the fermionic model, this follows from the polynomial dependence of the conjugated gates and the smooth action of the final Gaussian unitary.
In the bosonic model, the gates act smoothly on Schwartz space as in the proof of Lemma~\ref{lemma:bos-adaptive-kernel}, and the occupation bras define continuous linear functionals there.
Passing to the rays of the nonzero contractions therefore gives countably many smooth parameterizations with $O(M^2+tM)=O(m_{\mathrm{tot}}^2+tm_{\mathrm{tot}})$ real parameters each.

Comparing this count with the target dimensions in Secs.~\ref{appendix:fermionic-dimension-protocols} and~\ref{appendix:bos-dimension-protocols} gives the width-dependent lower bounds in \eqref{equation:mode-gate-tradeoff} when $m_{\mathrm{tot}}\le t$. When $m_{\mathrm{tot}}>t$, we have $tm_{\mathrm{tot}}>t^2$, so Theorems~\ref{theorem:fermionic-lower} and~\ref{theorem:bos-code-lower} give the same bounds. Taking a countable union of the exceptional sets over $m_{\mathrm{tot}}$ and $t$ gives a common full-measure target set.
\end{proof}

\end{document}